\documentclass[5p, number,sort&compress,times,fleqn]{elsarticle}

\usepackage{amsmath}
\usepackage{amssymb}
\usepackage{graphicx}
\usepackage{color}

\usepackage[normalem]{ulem}
\newcommand{\REMOVEDTEXTMODE}[1]{\textcolor{blue}{\sout{#1}}}
\newcommand{\REMOVED}[1]{
\ifmmode
  \text{\REMOVEDTEXTMODE{$#1$}}
\else
  \REMOVEDTEXTMODE{#1}
\fi}
\newcommand{\ADDEDTEXTMODE}[1]{\textcolor{blue}{#1}}
\newcommand{\ADDED}[1]{
\ifmmode
  \text{\ADDEDTEXTMODE{$#1$}}
\else
  \ADDEDTEXTMODE{#1}
\fi}

\newcommand{\trans}{{\mathsf{T}}}

\begin{document}

\begin{frontmatter}

\title{Adding quadric fillets to quador lattice structures} 

\author{Fehmi Cirak\corref{cor1}}  
\author{Malcolm Sabin}  

\cortext[cor1]{Corresponding author}

\address{Department of Engineering, University of Cambridge, Trumpington Street, Cambridge CB2 1PZ, UK}

\begin{abstract}
Gupta et al.~\cite{qd, qp} describe a very beautiful application of algebraic geometry to lattice structures composed of quadric of revolution (quador) implicit surfaces.  However, the shapes created have concave edges where the stubs meet, and such edges can be stress-raisers which can cause significant problems with, for instance, fatigue under cyclic loading. This note describes a way in which quadric fillets can be added to these models, thus relieving this problem while retaining their computational simplicity and efficiency.

\end{abstract}

\newpage


\end{frontmatter}

Throughout, we use an upper case letter to denote both a surface (e.g., a sphere~$S$ with the centre $(x_c, \, y_c, \, z_c)^\trans$ and radius~$r$) and the implicit function (i.e.,  $S(x,\, y, \, z) = (x-x_c)^2+(y-y_c)^2+(z-z_c)^2 - r^2$) which is zero on that surface.  No confusion should arise. Let $S$ denote the quadratic function 
with the zero set on the central sphere of a quador hub, increasing outward from the centre. 
\begin{figure}[h!]
	\centering 
	\includegraphics[scale=0.575]{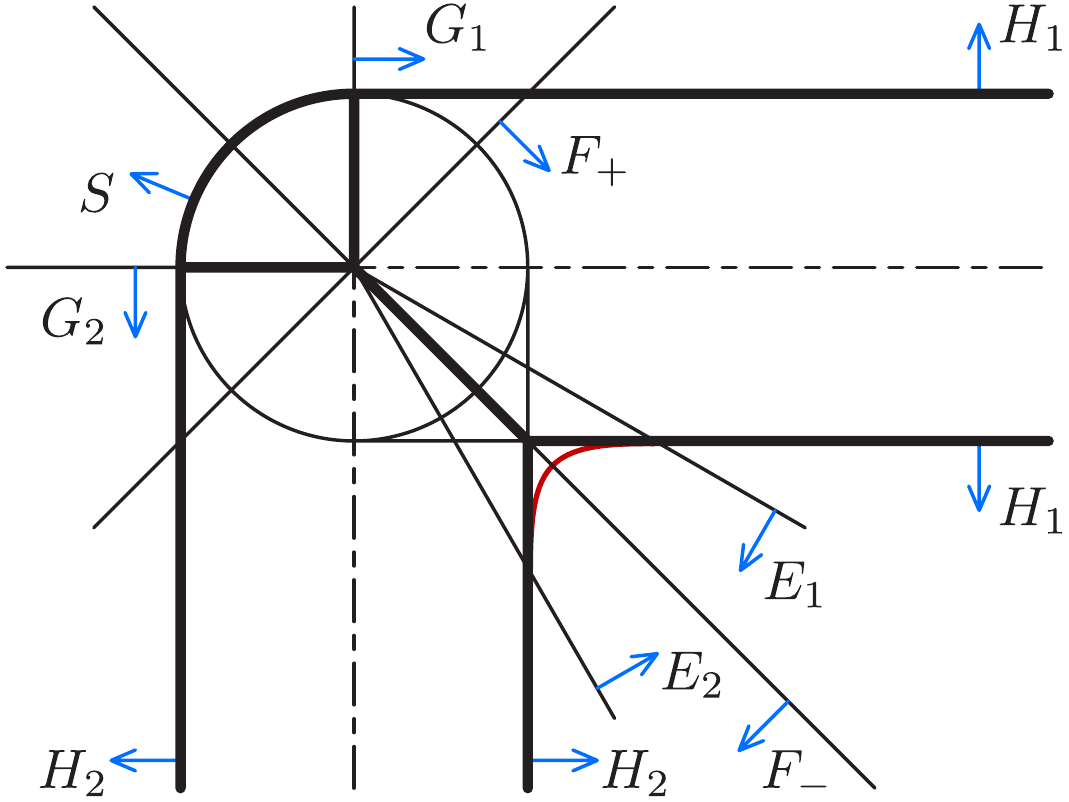}
  	\caption{A hub (lattice joint) with two attached stubs (beams leading to other joints).  This figure is diagrammatic only. Its purpose is to identify what surfaces the various letters denote. The arrow on a surface  indicates the gradient vector of the expression whose zero set is the surface.}
  \label{fig:introExample}
\end{figure}

Let $H_1$ and $H_2$ denote the quadratic functions of two quadors tangent to the central sphere~$S$.  Then, these two functions have the form
\begin{align*}
H_1  = S - G_1^2  \qquad &\text{ and }\qquad
H_2  = S - G_2^2
\end{align*} 
for linear functions $G_1$ and $G_2$ which are zero on the respective planes of tangency, so that~$\nabla H_1 = \nabla S$ and $\nabla H_2 = \nabla S$.

We can construct planes, for constants $ \alpha >0$ and $\beta>0$,
\begin{align*}
E_1= \alpha F_+ + \beta F_- \qquad&\text{ and }\qquad
E_2= \alpha F_+ - \beta F_- 
\end{align*}
with the functions $F_- = G_2-G_1$ and $F_+ = G_2+G_1$.

Consider the quadrics  whose equations are $H_1 -E_1^2=0$ (which is tangential to $H_1$ along the curve of intersection with $E_1$)  and $H_2-E_2^2=0$ (which is  tangential to $H_2$ along the curve of intersection  with $E_2$). These two will be the same quadric, providing a fillet between $H_1$ and $H_2$ if
\begin{align*}
  H_1-E_1^2 &= H_2-E_2^2  \\ 
  \text{or\qquad} H_1-H_2 &=  E_1^2-E_2^2 \\
  \text{but\qquad} H_1-H_2 &= S-G_1^2 - S+G_2^2 
  = G_2^2-G_1^2 \\ 
  &= (G_2+G_1)(G_2-G_1) \\
  &= F_+F_- \\
  \text {and\qquad }E_1^2-E_2^2 &= (E_1+E_2)(E_1-E_2) \\
  &= (2\alpha F_+)(2\beta F_-) \\
  &= 4\alpha\beta F_+F_- 
 \end{align*}
so we get a single fillet quadric if $\alpha\beta=1/4$.   We can choose either $\alpha$ or $\beta$ and then the other is fixed.   The ratio between the two controls the angles of the planes $E_1$ and $E_2$ either side of~$F_-$. Increasing $\beta$ slowly from zero increases the size of the fillet and its smallest radius of curvature, but this increases the length of the stub.

The above shows that there exists a fan of possible quadrics providing a tangent continuous join between adjacent stubs. Each piece of surface has an exact implicit form, and an exact parametric form.  The curves of tangency are all conics with exact parametric curves, exact implicit curves within their planes, and exact p-curves (i.e., trimming curves in parameter space) within both surfaces.  Surfaces can be separated by the planes which contain the curves of tangency, and so all the important properties in \cite{qd} and \cite{qp} still apply. 
\vspace{-0.3cm}
\bibliographystyle{elsarticle-num-names}

\end{document}